\def\final {1}			

\ifdefined\preprint
  \documentclass[preprint,review,12pt]{elsarticle}
\fi
\ifdefined\wordcount
  \documentclass[final,3p,times,twocolumn]{elsarticle}
\fi
\ifdefined\final
  \documentclass[final,3p,times,twocolumn]{elsarticle}
\fi

\usepackage{color}
\usepackage{multirow}
\usepackage{xcolor}
\def\wordcount{1}

\usepackage{nomencl}
\makenomenclature

\usepackage{amsmath}
\usepackage{siunitx}
\biboptions{sort&compress}

\journal{Combustion and flame}

\begin{document}

\begin{frontmatter}

\title{\textcolor{black}{Coupled interaction between acoustics and unsteady flame dynamics during the transition to thermoacoustic instability in a multi-element rocket combustor}}
\author[fir]{Praveen Kasthuri}
\author[fir]{Samadhan A. Pawar}
\address[fir]{Department of Aerospace Engineering, Indian Institute of Technology Madras, Chennai 600036, India.}
\author[sec]{Rohan Gejji}
\author[sec]{William Anderson}
\address[sec]{School of Aeronautics and Astronautics, Purdue University, West Lafayette, IN 47907, U.S.A.}
\author[fir]{\\ R. I. Sujith \corref{cor1}}
\ead{sujith@iitm.ac.in}

\cortext[cor1]{Corresponding author: R. I. Sujith}

\begin{abstract}
\textcolor{black}{Rocket engine combustors} are prone to transverse instabilities that are characterized by large amplitude high frequency oscillations in the acoustic pressure and the heat release rate. In this work, we study the coupled interaction between the acoustic pressure and \textcolor{black}{the CH* intensity (representative of heat release rate)} oscillations in a 2D multi-element self-excited model rocket combustor during the transition from a stable state to thermoacoustic instability through intermittency. We show the emergence of synchronization between these oscillations from desynchronization through intermittent phase synchronization during the onset of thermoacoustic instability. We find substantial evidence that the \textcolor{black}{intensities of} the jet flames close to the end wall is higher than that observed near the center of the combustor as a result of its strong coupling to the local acoustic field. Using concepts from recurrence theory, we distinguish the type of synchronization between the acoustic pressure and the \textcolor{black}{CH* intensity} oscillations at the end wall and the center of the combustor during thermoacoustic instability. Analyzing the local \textcolor{black}{CH* intensity} oscillations, we observe that the longitudinally propagating jet flames experience substantial transverse displacement with flame merging effects during thermoacoustic instability. Furthermore, we differentiate the interaction of the jet flames with the shock wave during intermittency and thermoacoustic instability at both the end wall and the center of the combustor. We also discerned the change from stochastic to deterministic nature in the local \textcolor{black}{CH* intensity} oscillations during the transition to thermoacoustic instability. Our results demonstrate that only the first few transverse modes contribute to the generation of acoustic power from the reacting flow, in spite of the pressure oscillations featuring several harmonics. The insights gained from such coupled analysis during the transition from stable state to thermoacoustic instability can be leveraged to aid modeling efforts of thermoacoustic instability in rocket engines with multiple injectors. 
\end{abstract}

\begin{keyword}
rocket engine \sep self-excited \sep transverse thermoacoustic instability \sep coupled interaction \sep intermittency \sep recurrence quantification analysis
\end{keyword}

\end{frontmatter}

\ifdefined \wordcount
\clearpage
\fi

\section{Introduction}
\label{introduction}
Transverse thermoacoustic instability continues to plague the development of stable rocket engine combustors which propel launch vehicles intended for earth orbits and outer space \cite{anderson1995liquid}. The large amplitude periodic pressure and heat release rate oscillations during transverse thermoacoustic instability can overwhelm the thermal protection mechanisms in the thrust chamber of rocket engines. 
If the chamber pressure oscillations exceeds a certain amplitude, it can lead to structural failure of the engine; followed either by a premature engine shutdown or a catastrophic explosion, both resulting in a mission failure \cite{anderson1995liquid}. Often, engineers devise control measures, empirically found during extended development tests, to arrest the growth of the amplitude of the oscillations. Hence, it is desirable to address combustion stability related design changes at an earlier stage of development in order to minimize the associated costs. 


It is widely accepted that thermoacoustic instability occurs when the heat release rate fluctuations are in-phase with the acoustic pressure oscillations inside the rocket combustor. This relationship is quantified by the Rayleigh criterion \cite{rayleigh1878explanation} to estimate the balance between the acoustic driving and the acoustic damping in a combustor. If the driving exceeds the damping, the oscillations grow in amplitude or vice versa. Several seminal theoretical studies have helped us to characterize the dynamics in the combustion chambers of rocket engines by understanding the growth of the oscillations, the occurrence of DC shift (i.e., rise in mean pressure levels), etc. \cite{flandro2007nonlinear,culick2006unsteady,sirignano2015driving}. \textcolor{black}{To understand the onset of thermoacoustic instability in multi-injector rocket combustors, it is essential to analyze the coupled behavior between the acoustics, various jet flames and flow processes. So far, several mechanisms have been reported to initiate and sustain high-frequency thermoacoustic instabilities in such rocket combustors. }

\textcolor{black}{While delivering propellants is the primary job of injectors, any perturbation developed upstream of the combustor (i.e., in the feedlines, turbopumps etc.) can interact with the dynamics in the combustion chamber via the injector. When a feedback loop is established between the injector resonant modes and the chamber acoustics, an undesirable growth in the amplitude of pressure oscillations leading to thermoacoustic instability can materialize \cite{bazarov1998liquid,groning2016injector}. } 


\textcolor{black}{Based on the location at which the highest heat release rate oscillations are recorded during the occurrence of thermoacoustic instability, the instability sustaining mechanism could be velocity coupled or pressure coupled with the resulting heat release rate oscillations. In the velocity coupled mechanism, the jet flames near the acoustic velocity antinode are perturbed more than the jet flames at other locations \cite{rey2004high}. These perturbations leads to a nonuniform distribution of the vortices across the combustor. Eventually, a collective interaction between the neighboring jet flames results in high heat release rate oscillations leading to thermoacoustic instability. In the pressure coupled mechanism, the largest heat release rate oscillations are observed at the acoustic pressure antinodes \cite{knapp2007interaction}. Morgan et al. \cite{morgan2015comparative} used dynamical mode decomposition obtained from chemiluminescence images to show the presence of velocity coupling for the first transverse mode and a pressure coupling for the second transverse mode near the center of a 2D rocket combustor. } 

\textcolor{black}{Richecoeur et al. \cite{richecoeur2006high} showed that flame-flame interaction could lead to the establishment of thermoacoustic instability. They reported that at certain operating conditions, the shear regions of neighboring jet flames collide to enhance the atomization and mixing of propellants, leading to an increase in the intensity of heat release rate oscillations. Recently, M{\'e}ry \cite{mery2017impact} demonstrated that the transverse flame displacement significantly affects the heat release rate oscillations in a model rocket combustor under external excitation. Apart from these aforementioned mechanisms, several other mechanisms pertaining to atomization and vaporization of liquid propellants have also been reported to initiate and sustain high frequency thermoacoustic instability in the combustor of rocket engines \cite{anderson1995liquid,harrje1972liquid}. }

\textcolor{black}{Recently, the use of high-fidelity simulations have enhanced our understanding of the mechanisms driving high-frequency thermoacoustic oscillations in rocket engine combustors. Performing large eddy simulations (LES) of multiple cryogenic jet flames interacting with high frequency transverse acoustic modes, Hakim et al. \cite{hakim2015large} reported that the location of the jet flame in the transverse acoustic environment considerably influences the unsteady heat release rate oscillations. Urbano et al. \cite{urbano2016exploration} performed LES on a LOx/H2 rocket engine featuring 42 coaxial injectors. They identified that the heat release dynamics is dependent on the relative position of the flame with respect to the acoustic pressure and velocity nodes. Weakened flame response was observed at the pressure nodes suggesting that the lateral motion of the jet flames imparted by the transverse velocity perturbations do not provide enough energy to sustain thermoacoustic instability \cite{urbano2017study}. Harvazinski et al. \cite{harvazinski2015coupling} performed hybrid RANS/LES of a model rocket combustor exhibiting self-excited longitudinal instabilities. They identified that relative phase between the acoustic pulses in the oxidizer tube of the shear coaxial injector and the combustion chamber results in either stable, marginally stable or unstable behavior.}

Despite decades of active research, a deeper understanding of the coupled interaction between the acoustic and the heat release rate oscillations in rocket engine combustors has eluded us \cite{harrje1972liquid,hardi2012experimental,sardeshmukh2019high}. Moreover, past studies have focused mostly on the state of thermoacoustic instability. Hence, the coupled interaction between the acoustic pressure and the heat release rate oscillations in the combustors of rocket engines during the transition from stable state to thermoacoustic instability is not completely understood. 
The nonlinear interactions between acoustics, multiple jet flames, and flow processes in rocket engine combustors introduce additional features such as steep fronted waves \cite{saenger1960periodic} and rise in mean pressure \cite{flandro2007nonlinear}.
Further, the acoustically non-compact nature (i.e., the flame and the acoustic length scales are comparable) due to high frequency acoustics typically seen in combustors of rocket engines, implies that the acoustic perturbations vary significantly across the flame. This complicates the study of the coupled interaction between the acoustics and heat release rate in the combustor. 

In this study, we examine measurements from experiments performed on a self-excited multi-element subscale transverse rocket engine combustor operating on an oxidizer-rich staged combustion cycle. 
Recently, Kasthuri et al. \cite{kasthuri2019dynamical} showed that the transition from stable state to thermoacoustic instability in this combustor occurs via intermittency. Here, we adopt the framework of synchronization theory \cite{lakshmanan2011dynamics,pawar2017thermoacoustic} and recurrence theory \cite{marwan2007recurrence,gotoda2014detection} to analyze the coupled dynamics of acoustic pressure and heat release rate oscillations during this transition. This work is inspired from the recent progress in thermoacoustics based on a complex systems approach where the system is treated as a whole instead of a sum of its components \cite{sujith2021dynamical}. \textcolor{black}{We embrace this complex systems approach and understand the coupled interaction between the acoustic pressure and the CH* intensity oscillations in the presence of turbulent flow and shock wave in the combustor.}


First, we perform a temporal analysis of the coupling between the acoustic pressure and the CH* intensity (representative of heat release rate) oscillations during the transition to thermoacoustic instability. Then, we study the spatiotemporal dynamics of the jet flames near the end wall and the center of the combustor for each dynamical state. We also estimate the contribution of each transverse mode to the generation of acoustic power using the spatial distribution of Rayleigh index. Using recurrence quantification analysis, we quantify the extent of determinism in the dynamics of local CH* intensity oscillations at both the end wall and center locations of the combustor.


\section{Experimental configuration}
\begin{figure*}[t!]
\centering
\includegraphics[scale = 0.7] {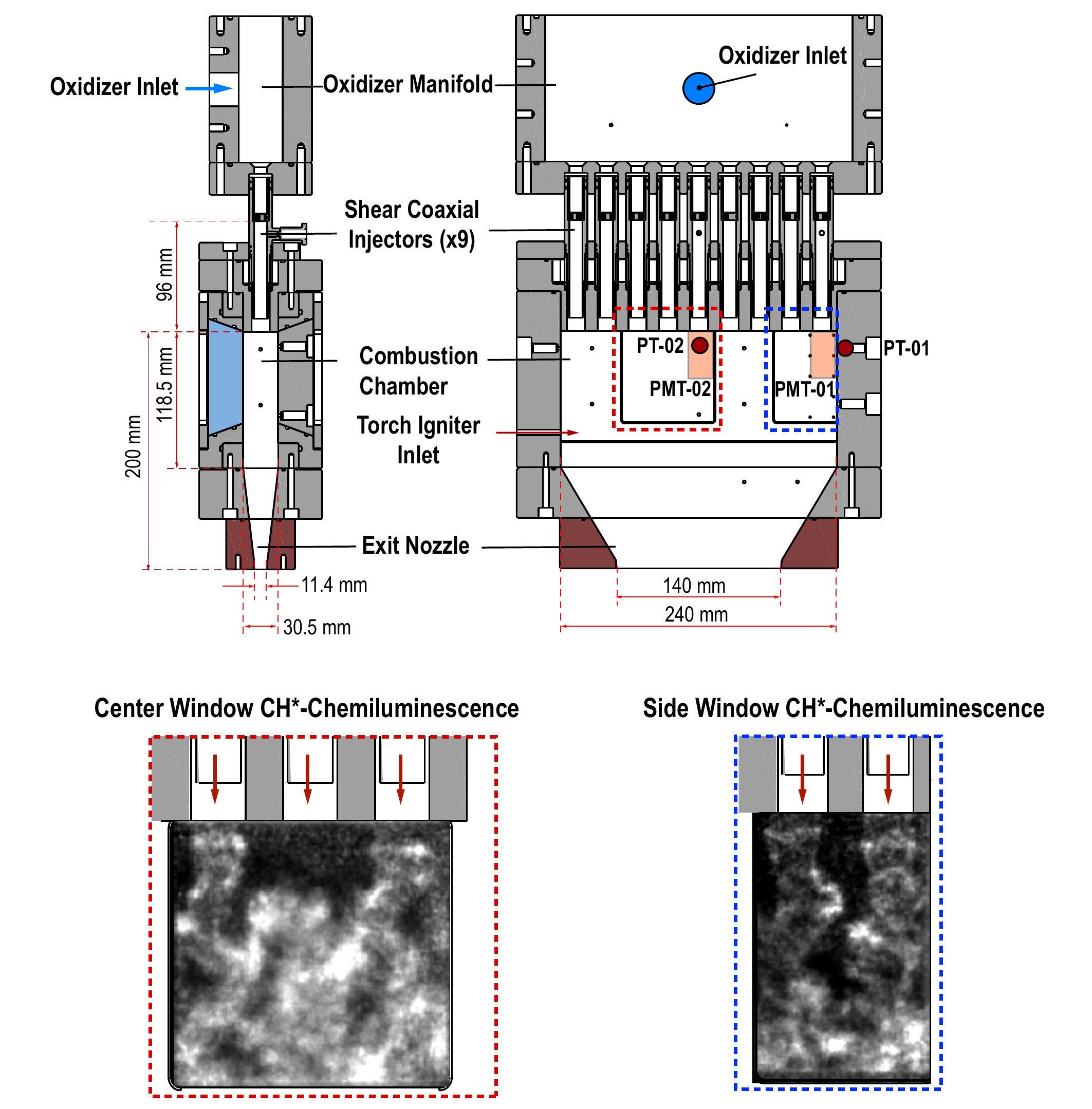}
\caption{Schematic of the multi-element rocket engine combustor without the preburner.  Representative flame images as observed from the optically accessible windows located near the center and the end wall of the combustor are shown. \textcolor{black}{The high-frequency pressure transducer and photomultiplier tube measurement locations are labeled as PT and PMT, respectively.} }
\label{r0}
\end{figure*}

The experiments are performed in a multi-element self-excited subscale rocket combustor based on an oxidizer-rich staged combustion cycle (Fig.~\ref{r0}). This combustor is an evolution of combustors designed to excite transverse mode instabilities \cite{orth2018measurement}. The reactant mixture comprises of gaseous oxygen preheated to 620 K in a hydrogen-fed preburner, and methane (at 297 K). The combustor houses a linear array of nine oxidizer centered gas-gas shear coaxial injectors located at the entry to the combustion chamber. An oxidizer manifold feeds the hot oxidizer (oxygen with 4 - 5\% mass fraction of water vapor) uniformly to each of the coaxial injectors. Methane is injected through each of the shear coaxial injectors at the downstream end of the oxidizer posts through a manifold with a choked inlet. In turn, nine turbulent jet flames are established in the combustor. The propellant flow rates were chosen to achieve an equivalence ratio of approximately 1.24, which is typical for liquid rocket engines. The mean Mach number in the oxidizer posts is 0.265 at nominal operating conditions. A mean pressure level of $\sim$1.14 MPa is maintained over the course of a test. All the injectors are uniformly spaced out by an injector centerline-centerline distance of 25.7 mm. The exit diameter of the injector element is 15.7 mm. The combustor walls are coated with a protective layer of thermal barrier coating to minimize the wall heat loss during the test interval. The cumulative mass flow rate of oxidizer is 0.71 kg/s, while the mass flow rate of methane is maintained at 0.22 kg/s. Propellant flows were metered using critical flow venturi nozzles upstream of the choked inlets to the propellant manifolds \cite{ASMEMFCCommittee2016,ISO93002005}. Uncertainty of mass flow rates, and subsequently operating conditions, were evaluated using the Kline-McClintock method of uncertainty propagation \cite{Kline1953} and following the procedure presented by Walters et al. \cite{Walters2020}. The typical uncertainty in mass flow rate of propellants was $\leq$1\% with a 95\% confidence interval. 

The geometry and operating conditions are devised such that only transverse modes (1T mode frequency at 2650 Hz) are excited in the combustor. The combustion chamber is 240 mm wide and 30.5 mm deep. The length of the combustion chamber is 200 mm and is split into a 118.5 mm straight section and 81.5 mm converging section. The combustion chamber is terminated with a converging section designed to preempt longitudinal modes and ensure acoustic decoupling from the downstream locations. The length of the chamber is designed for a fundamental longitudinal (1L) mode of 3475 Hz. This choice of length ensures that the harmonics of the transverse modes do not match with the 1L mode or its harmonics. A nozzle at the end of the converging section provides a choked boundary condition at the exit of the combustor. Ignition of preburner and main combustion chamber is achieved using hydrogen-oxygen torch igniters. A test run providing approximately 1 s of steady inflow provides sufficient time to acquire data exhibiting transition to thermoacoustic instability at the frequencies of interest with negligible effects of heat loss \cite{orth2017experimental}. A summary of operating conditions for the representative test cases discussed in the current work are provided in Table.~\ref{tab:op}.

\begin{table}[h!]
 \caption{Summary of operating conditions}
 \centering
 \begin{tabular}{cccccccccc}
 \hline
 $\dot{m}_{Ox}$ & $\dot{m}_{fuel}$ & $\phi$ & $P_{c}$ & $T_{Ox}$ & $T_{fuel}$ & $f_{1T}$\\\
 [kg/s] & [kg/s] & [-] & [MPa] & [K] & [K] & [kHz]\\\hline
 0.71 & 0.22 & 1.24 & 1.14 & 620 & 297 & 2.65\\
 \hline
 \end{tabular}

\label{tab:op}
 \end{table}

In this study, we analyze two spatial regions of interest separately through tests - A and B performed under the same set of operating conditions. For test A, the optically accessible region is located towards the right end of the combustor, while for test B, this region is located near the center. Thus, we observe two jet flames for test A and three jet flames for test B (see Fig.~\ref{r0}). We acquire the acoustic pressure oscillations at the right-side end wall \textcolor{black}{(PT-01)} and the center of the combustor \textcolor{black}{(PT-02)} using piezoresistive Kulite WCT-312M sensors, at a rate of 250 kHz. The pressure sensors are installed in a recessed Helmholtz cavity to provide thermal isolation from the combustion gases. The resonance frequency of the cavity is designed to be higher than any frequencies of interest in the experiment ($\sim$22 kHz) \cite{FuggerRSI2017}. This installation enables accurate measurement of dynamic pressure fluctuations in the chamber while reducing thermal load on the sensor element. 

\textcolor{black}{Based on simulations and experiments performed for similar pressures and operating conditions, Bedard et al. \cite{bedard2020chemiluminescence} and Sardeshmukh et al. \cite{sardeshmukh2017use} compared heat release rate to chemiluminescence from CH*, OH* and CO2* radicals in the flame. They concluded that the CH* chemiluminescence provided a better qualitative representation of the heat release rate dynamics even though a phase difference was reported between the experimentally obtained CH* emissions and the heat release obtained from the computations.}

A Hamamatsu photomultiplier tube module (H11903-210) attached to a fiber optic probe gathered line of sight light emissions from a volume in the optically accessible window \textcolor{black}{(PMT-01 and PMT-02)} accessible in the combustor. The light emissions are filtered using an optical filter (Semrock FF01-427/10) to obtain CH* chemiluminescence signals at the same rate of 250 kHz synchronous with the acoustic pressure measurements \cite{bedard2017detailed}.

\textcolor{black}{Line-of-sight integrated} high speed CH* chemiluminescence images are simultaneously recorded at a rate of 100 kHz through the optically accessible windows in the combustion chamber. An optical filter (Semrock 434/14 Brightline Bandpass) of 14 nm bandwidth centered at 434 nm isolated the CH* emissions from the background luminosity. \textcolor{black}{The emissions are collected through a 200 mm focal length, $f$/4.0 objective (Nikon AF Micro NIKKOR) and then amplified by a Lambert HiCATT 25 intensifier with 1:1 relay lens, and recorded with a Phantom v2512 high speed CMOS camera with a spatial resolution of 0.214 mm/pixel. The intensifier gain was set at 750 V with an exposure of 1 $\mu$s. \textcolor{black}{The same camera and intensifier settings were used for both the tests.}} 

\textcolor{black}{The CH* chemiluminescence measurements from the high speed imaging are representative of the local heat release rate dynamics. However, the photomultiplier measurement is representative of the cumulative heat release rate measurement since the obtained emissions emanates from a probe volume rather than just a point. \textcolor{black}{In the rest of the paper, we refer to the photomultiplier measurement and the high speed CH* chemiluminescence images as CH* intensity oscillations and local CH* intensity oscillations, respectively.}} Further details of the geometry of the experimental rig, the operating conditions, hardware and the measurement techniques can be found in Orth et al. \cite{orth2018measurement}. 

\section{Results and discussions}

\textcolor{black}{In this paper, we study the coupled interaction between the turbulent reacting flow field and acoustic field during the transition to transverse thermoacoustic instability. We define the acoustic field and the turbulent reacting field as the two oscillators and use the framework of synchronization theory to understand the interacting between these two subsystems.}

\textcolor{black}{For the study of synchronization, the oscillators should be self-sustained and exhibit a coupling (weak/strong) between them. In turbulent combustors, due to the inherent hydrodynamic fluctuations, the heat release rate and the acoustic pressure exhibit self-sustained chaotic oscillations during stable operation of the combustor \cite{pawar2017thermoacoustic}. The acoustic field in a cold turbulent flow has been reported to exhibit self-sustained chaotic oscillations at broadband frequencies \cite{godavarthi2018coupled}. In our study, the underlying flow field is highly turbulent (Reynolds number $\sim$ 393000 at the injector exit), and as the oxidizer is preheated to 635 K, the jet flames inherently oscillate even during stable operation due to density-stratification \cite{emerson2012density} and also due to inherent turbulent fluctuations. Therefore, in the presence of turbulent reactive flow, the acoustic field and the turbulent reacting flow in our combustor can indeed be considered as self-sustained aperiodic oscillators.}

\textcolor{black}{Note that each of these oscillators behave as a damped oscillator in the absence of turbulent flow. However, the presence of continuous disturbances from the inherent turbulent hydrodynamic flow makes them self-sustained oscillators. Therefore, we can apply the framework of synchronization to study the coupled behavior between acoustic pressure and CH* intensity (representative of heat release rate) oscillations during the transition to thermoacoustic instability. Studying such synchronization behaviors between different subsystems (through different variables that represent these subsystems) of the same system is well-established in nonlinear dynamics; examples include thermoacoustic systems \cite{pawar2017thermoacoustic,chiocchini2018chaotic,murayama2019attenuation,guan2019chaos}, biological systems \cite{schafer1999synchronization}, psychology \cite{scherer_2000}, neuroscience \cite{siapas2005prefrontal,nikulin2006phase}, and network systems \cite{pecora2015synchronization}.}


Here, we study the data obtained from two test cases - A and B. Since both the tests are performed for the same set of operating conditions, we will utilize test A and test B for comparing the dynamics near the end wall and the center of the combustor, respectively. First, we will investigate the temporal coupled behavior between the acoustic pressure and the \textcolor{black}{CH* intensity} oscillations during the onset of thermoacoustic instability using measures from synchronization theory. Subsequently, using the methods of phase averaging and recurrence quantification, we will perform a spatial analysis of the \textcolor{black}{local CH* intensity oscillations in the reaction zone} and analyze the spatial distribution of acoustic power sources/sinks for each dynamical state exhibited by the combustor. Throughout the rest of the paper, overline and prime are used to denote mean and fluctuations, respectively. 

\subsection{Temporal analysis of the coupled acoustic pressure and \textcolor{black}{CH* intensity} fluctuations}
\begin{figure*}[t!]
\centering
\includegraphics[scale = 0.072] {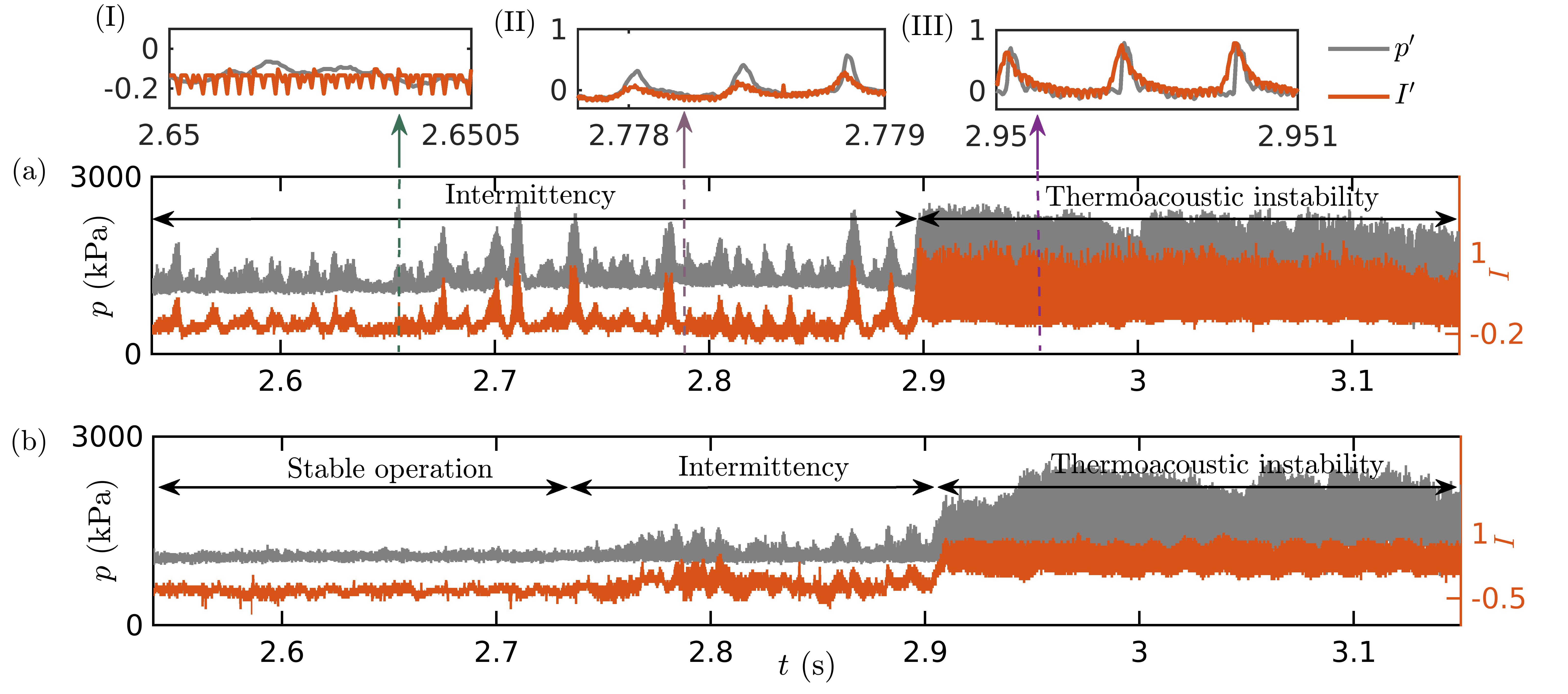}
\caption{Time series of the pressure ($p$) and the normalized \textcolor{black}{CH* intensity ($I$)} oscillations for (a) test A and (b) test B. The zoomed insets represent the normalized waveforms of $p'$ and \textcolor{black}{$I'$} during (I) aperiodic epoch of intermittency, (II) periodic epoch of intermittency, and (III) thermoacoustic instability. }
\label{r1}
\end{figure*}

In Fig.~\ref{r1}, we show the pressure (i.e., $p$ = $\bar{p}+p'$) and \textcolor{black}{the normalized CH* intensity ($I$)} measured at both the locations in the combustor during the transition to thermoacoustic instability. The pressure and photomultiplier signals acquired near the end wall \textcolor{black}{(PT-01 and PMT-01)} and the center of the combustor \textcolor{black}{(PT-02 and PMT-02)} are used to capture this dynamical transition in tests - A and B, respectively. To visually compare the qualitative behavior of the two signals, we show the waveforms of the acoustic pressure ($p'$) and the \textcolor{black}{CH* intensity fluctuations ($I'$)} normalized with their respective maximum values in the zoomed insets. 

Although both the tests are performed for the same set of operating conditions, interestingly, they exhibit different sets of dynamical states. In test A, we observe intermittency, characterized by epochs of low amplitude aperiodic oscillations ($p'/\bar{p}\sim$10\%) and high amplitude periodic oscillations ($p'/\bar{p}\sim$90\%) interspersed in a random manner. This intermittency is succeeded by thermoacoustic instability, characterized by large amplitude ($p'/\bar{p}\sim$100\%) high frequency ($\sim$2650 Hz) limit cycle oscillations at the first transverse mode (1T) of the combustor. At the beginning of test B, we observe a stable state, characterized by low amplitude aperiodic oscillations ($p'/\bar{p}<$10\%). This stable state is followed by intermittency and thermoacoustic instability. The transition to thermoacoustic instability takes places without any change in the control parameter in both the tests. A detailed characterization of the temporal variation of the acoustic pressure oscillations for these cases are discussed in \cite{kasthuri2019dynamical}. 

Next, we qualitatively analyze the coupled behavior of the $p'$ and \textcolor{black}{$I'$} during the onset of thermoacoustic instability. At the outset, we observe that the oscillations are either aperiodic or periodic during each dynamical state en route to thermoacoustic instability. During thermoacoustic instability (Fig.~\ref{r1}III), we observe that both $p'$ and \textcolor{black}{$I'$} are periodic and nearly in-phase with each other.
The $p'$ signal exhibits a spiky wave form where the amplitude of the signal rises sharply to a high value, then decays gradually and stays near the minimum amplitude for a long epoch in the oscillation cycle \cite{kasthuri2020recurrence}. Such a waveform is typical of steep fronted pressure wave. When the compression front of the wave catches up with its expansion front, it leads to the formation of steep fronted waves \cite{saenger1960periodic,kasthuri2019dynamical}. While the waveform of \textcolor{black}{$I'$} is also spiky and stays near the minima for a long epoch in the oscillation cycle, the rise in its amplitude is not as rapid as that of the $p'$.

During intermittency, we observe alternate occurrences of bursts of periodic oscillations in between epochs of low amplitude aperiodic oscillations. The wave steepening effect is not as pronounced during the periodic part of intermittency when compared to thermoacoustic instability (compare zoomed insets - II and III in Fig.~\ref{r1}). During the periodic part of intermittency (Fig.~\ref{r1}II), the oscillations of both $p'$ and \textcolor{black}{$I'$} match in their rhythms with nearly zero phase difference. Conversely, during epochs of aperiodic oscillations in  the intermittency signal (Fig.~\ref{r1}I), the temporal locking of oscillations in $p'$ and \textcolor{black}{$I'$}  appears to be absent. The waveform of the oscillations in each dynamical state of $p'$ and \textcolor{black}{$I'$} during test B is nearly akin to that seen in test A. This test exhibits the stable state before intermittency. During the stable state, we observe sustained desynchronized aperiodic oscillations in both $p'$ and \textcolor{black}{$I'$} signals.

\subsubsection{Cross wavelet analysis of the acoustic pressure and the CH* intensity oscillations}
To understand the synchronization characteristics between $p'$ and \textcolor{black}{$I'$} signals during the onset of thermoacoustic instability quantitatively and also to identify the locking of their dominant modes, we perform a cross wavelet transform (XWT) \cite{grinsted2004application} between these signals. The XWT indicates the regions in the time-frequency space where the two time series simultaneously exhibit high spectral powers. The complex Morlet wavelet shown in Eq.~(\ref{eqs1}) is used as the mother wavelet ($\psi_0$). Here, $\eta$ and $\omega$ are dimensionless time and frequency, respectively. The Morlet wavelet, $\psi_0$, can be dilated and translated in time-frequency space following a continuous wavelet transform \cite{grinsted2004application}.
\begin{align}
\begin{split}
\psi_{0}(\eta) &= \pi^{-1 / 4} e^{i \omega_{o} \eta} e^{(-1 / 2) \eta^{2}},
\\
W(u, s) &= \frac{1}{\sqrt{s}} \int_{-\infty}^{+\infty} x(t) \psi_0\left(\frac{t-u}{s}\right) \mathrm{d} t.
\end{split} 
\label{eqs1}
\end{align} 
Here, $x(t)$ is the time series analyzed, whereas $s$ and $u$ are the scale and translation parameters of $\psi_0$, respectively. The XWT for the two time series, $x(t)$ and $y(t)$ is obtained by $W_{xy}=W_{x} W_{y}^*$ (* denotes complex conjugate). The power spectrum is obtained as $|W_{xy}|$. The instantaneous relative phase between the two time series is given by $\arg(W_{xy})$. This information can be illustrated in a time-frequency plot.

\begin{figure*}[t]
\includegraphics[scale = 0.072] {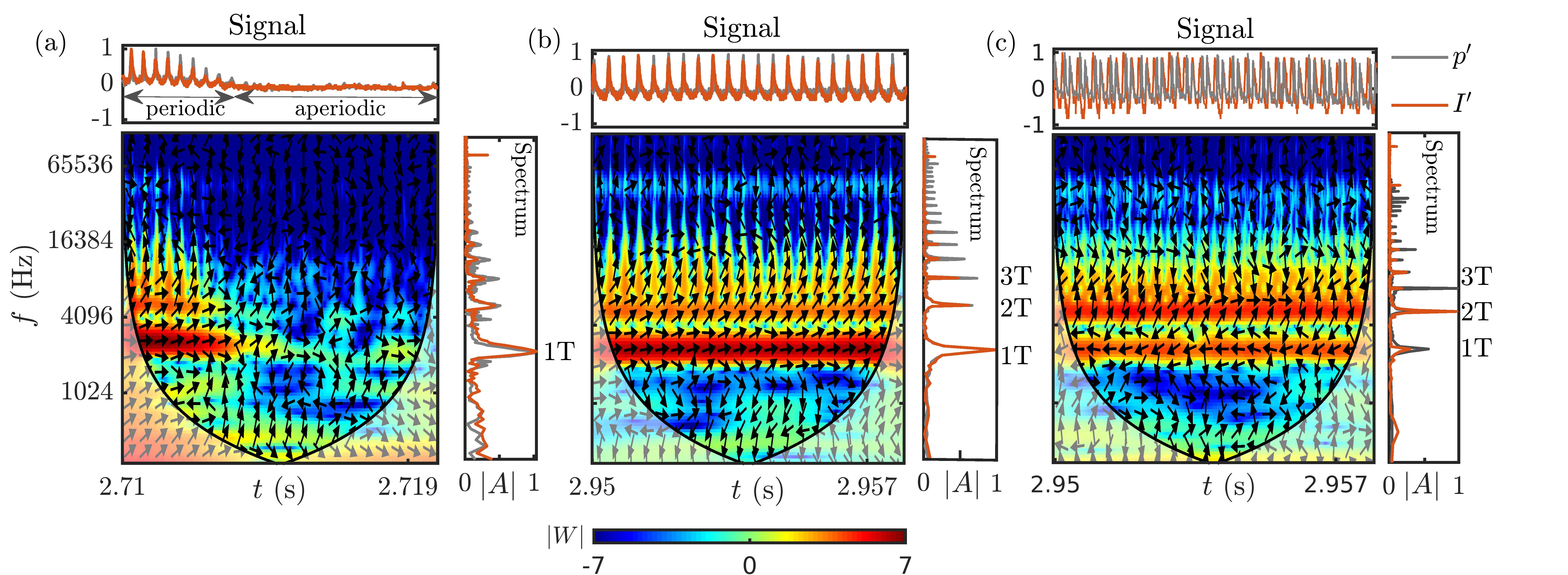}
\caption{The normalized time series of the acoustic pressure oscillations ($p'$) and \textcolor{black}{the CH* intensity oscillations ({$I'$})} for (a) periodic to aperiodic transition observed during intermittency, and (b) thermoacoustic instability measured near the end wall, and (c) thermoacoustic instability near the center of the combustor. We also show the corresponding XWT and the normalized amplitude spectrum from FFT. \textcolor{black}{PT-01 and PMT-01 are used in (a,b), while PT-02 and PMT-02 are used in (c).} The orientation of the arrows in the XWT represent the relative phase angle between $p'$ and \textcolor{black}{$I'$}. \textcolor{black}{A statistically sufficient time interval of around 21 oscillation cycles is used to evaluate the XWT plots. Note that the frequency scale used is nonlinear.}}
\label{r2}
\end{figure*}

In Fig.~\ref{r2}a, we show the XWT of $p'$ and \textcolor{black}{$I'$} signals for an epoch of intermittency comprising the transition from periodic to aperiodic oscillations. The region outside the shaded area in the XWT is called the 'cone of influence'. In the shaded region, the estimates of power of cross wavelet transform between the $p'$ and \textcolor{black}{$I'$} signals cannot be ascertained above a 95\% confidence level, due to the finiteness of the data and temporal sampling \cite{grinsted2004application}. Henceforth, we utilize the information inside the 'cone of influence' to understand the time-frequency behavior. When the signal is periodic, we observe a strong amplitude content at a band of frequencies centered around 2650 Hz (1T mode) in the plot of XWT. The common spectral power of this mode decays as soon as the oscillations in the signal become aperiodic. In the aperiodic epoch of the intermittency signal acquired at the end wall, we observe the absence of any common power between the $p'$ and \textcolor{black}{$I'$} oscillations. 

On the other hand, during the state of thermoacoustic instability (Fig.~\ref{r2}b,c), we observe that a common band of frequencies around 1T mode sustains their high magnitude throughout the signal. Moreover, the presence of steepened shock wave results in the occurrence of several harmonics of considerable amplitudes. During thermoacoustic instability, near the end wall (see Fig.~\ref{r2}b), we observe that the common spectral power gradually decreases from 1T to 10T modes. However, for thermoacoustic instability, near the center of the combustor (see Fig.~\ref{r2}c), we only observe high common spectral power for the 2T mode followed by the 1T mode. Performing a simulation based on the same combustor, Harvazinski et al. \cite{harvazinski2019modeling} had reported that the pressure at the center of the chamber has a frequency that is double the frequency observed at the end wall. The results from XWT plots in Fig.~\ref{r2}b,c are in agreement with their study. Further, our result indicates that only the first two modes dominate the coupled behavior of $p'$ and \textcolor{black}{$I'$} in the center of the combustor. 

Additionally, the instantaneous phase difference between $p'$ and \textcolor{black}{$I'$} signals at each frequency is represented by arrows distributed all over the plot of XWT. The alignment of these arrows in the same direction in time, \textcolor{black}{for a frequency having high common spectral power indicates the presence of synchrony between the signals at that frequency}. Further, the orientation of arrows indicates the value of the relative phase between these synchronized oscillations at the frequency they are locked and enables us to identify lead-lag behavior. During thermoacoustic instability (Fig.~\ref{r2}b,c), the arrows are aligned at the same angle at the frequencies corresponding to the first few transverse modes. Near the end wall (Fig.~\ref{r2}b), during thermoacoustic instability, the arrows are nearly horizontal and pointing rightward for the 1T mode. This suggests the presence of in-phase synchronization between the $p'$ and \textcolor{black}{$I'$} signals with a relative phase difference around \ang{-6}. 

Near the center of the combustor (Fig.~\ref{r2}c), during thermoacoustic instability, we observe that the arrows are almost horizontal but point leftwards at the frequency corresponding to 1T mode. This indicates that the $p'$ at the center of the combustor and the \textcolor{black}{$I'$} are anti-phase synchronized with a relative phase difference of \ang{170}. Near the center of the combustor, we observe that the 2T mode is the most dominant followed by the 1T mode in the spectrum of \textcolor{black}{$I'$} (see spectrum in Fig.~\ref{r2}c). In the corresponding XWT plot, the arrows corresponding to the 2T mode are aligned at \ang{45}. This indicates that \textcolor{black}{$I'$} near the center of the combustor leads $p'$ and is driven by both the acoustic velocity and pressure oscillations. \textcolor{black}{The type of synchronization existing between $p'$ and $I'$ at both the locations in the combustor during thermoacoustic instability is detected using recurrence measures in Sec.~\ref{Sec:rec-tai}.}

In Fig.~\ref{r2}a, during periodic epochs of intermittency, the arrows are aligned at the same angle for the 1T mode. However, during the aperiodic epoch of intermittency (Fig.~\ref{r2}a), the arrows are almost randomly oriented in all directions, indicating desynchrony between $p'$ and \textcolor{black}{$I'$} signals. Thus, both $p'$ and \textcolor{black}{$I'$} signals are phase synchronized during thermoacoustic instability and are intermittently phase synchronized during intermittency. During stable state, we observe sustained desynchronized behavior similar to that observed during the aperiodic epochs of intermittency \textcolor{black}{(refer to Appendix A)}. A similar synchronization transition to longitudinal thermoacoustic instability in a gas turbine type combustor was reported in \cite{pawar2018temporal}.

Further, the amplitude spectra of $p'$ and \textcolor{black}{$I'$} (shown on the right side of the each XWT in Fig.~\ref{r2}b,c) during thermoacoustic instability show the presence of several harmonics (up to 10T). However, the XWT of these signals (Fig.~\ref{r2}b,c) indicates that the spectral power in the common frequency bands gradually diminishes beyond the first few harmonics of the $p'$ and \textcolor{black}{$I'$} signals.

\begin{figure*}[t!]
\centering
\includegraphics[scale = 0.153] {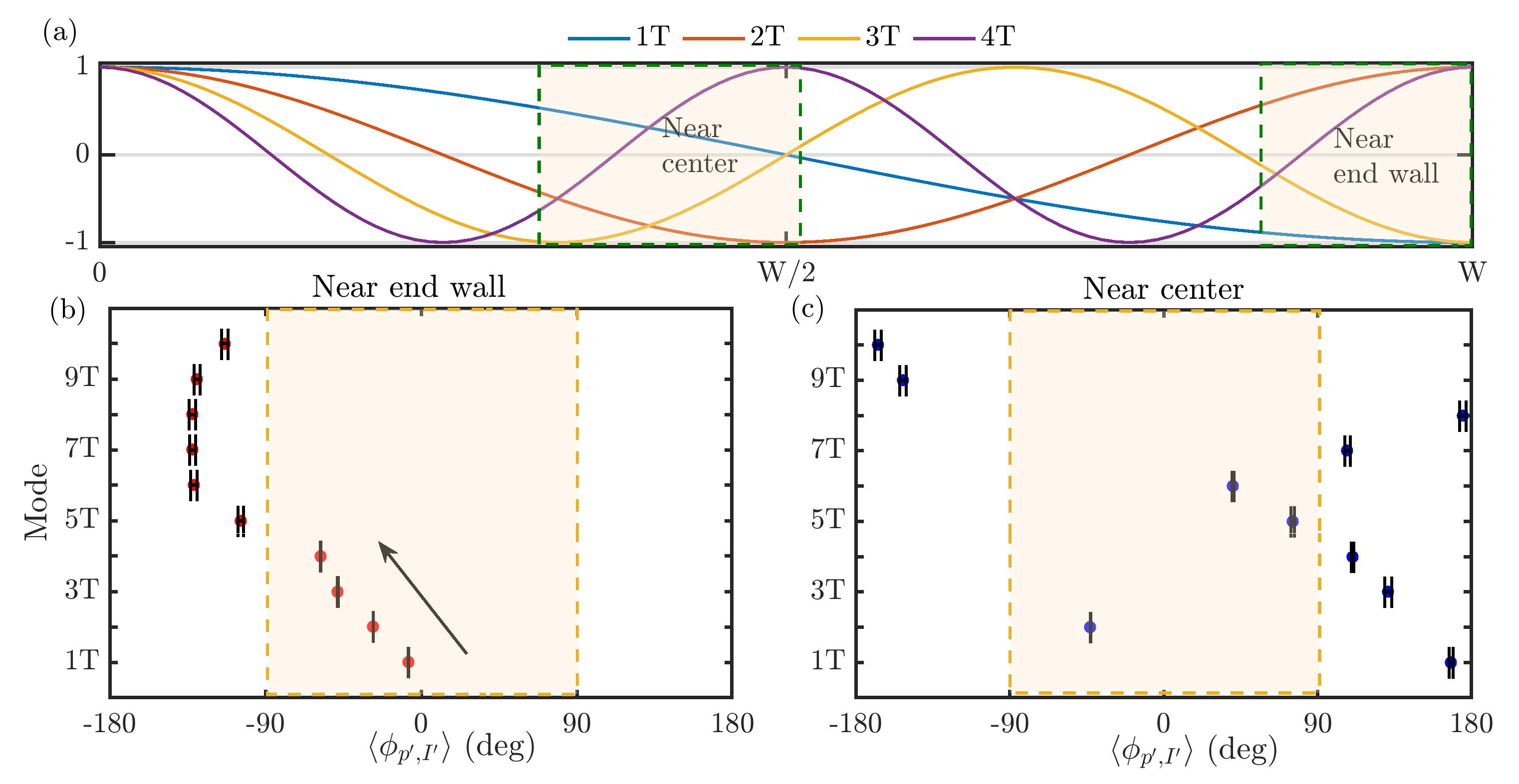}
\caption{(a) The mode shapes for the first four transverse acoustic \textcolor{black}{pressure} modes in the combustor. The mode shapes are derived from a cosine approximation. The mean relative phase ($\langle\phi\rangle_{p',I'}$) between the acoustic pressure and the \textcolor{black}{CH* intensity} oscillations for each transverse mode observed near the (b) end wall and (c) center of the combustor. The standard deviation of $\langle\phi\rangle_{p',I'}$ is captured by the span of the horizontal error bars.}
\label{r2b}
\end{figure*}

We plot the approximate mode shapes for the first four transverse acoustic \textcolor{black}{pressure} modes in Fig.~\ref{r2b}a. For the $n^{th}$ mode, its mode shape is calculated as $\cos(2\pi\frac{x}{W}n)$ with $W$ being the width of the combustor. \textcolor{black}{The presence of temperature and density gradients, shock wave, reactants and product species would significantly alter the mode shapes \cite{sujith1995exact}. In spite of these deficiencies, some features of the acoustic pressure mode shapes remain unchanged from Fig.~\ref{r2b}a. The end wall region houses the acoustic pressure anti-node for all the transverse modes. Further, the inherent flow and geometric symmetry of the combustor will ensure transverse symmetry in temperature. Therefore, the location of the pressure node for the first transverse mode will remain approximately at the center of the combustor.}

Using the XWT, we plot the variation of the mean of the relative phase ($\langle\phi\rangle_{p',I'}$) between $p'$ and \textcolor{black}{$I'$} obtained near the end wall (Fig.~\ref{r2b}b) and center (Fig.~\ref{r2b}c) of the combustor for each of the first ten transverse modes. Close to the end wall, $\langle\phi\rangle_{p',I'}$ is close to zero (albeit with the \textcolor{black}{$I'$} leading $p'$ slightly) for the 1T mode indicating the presence of strong coupling. However, beyond 1T mode, we observe a gradual change in $\langle\phi\rangle_{p',I'}$ from \textcolor{black}{\ang{0} to beyond \ang{90}}, denoting weaker coupling. For 2T to 4T modes, we observe that the \textcolor{black}{$I'$} leads $p'$. When the angle exceeds \ang{-90} (upward oriented arrows), we notice the absence of common power between these signals at those modes (see blue regions in Fig.~\ref{r2}b). This suggests that the two signals are desynchronized for harmonics greater than the 4T mode (see Fig.~\ref{r2b}a). 

The coupling behavior between $p'$ and \textcolor{black}{$I'$} at the center of the combustor is different from that near the end wall, due to the presence of the pressure node at the center for the 1T mode (which is the dominant mode near end wall). We observe strongest coupling for the 2T mode for which $\langle\phi\rangle_{p',I'}$ is around \ang{45} (Fig.~\ref{r2b}c). High power in the 2T mode is attributed to the center of the combustor being a pressure antinode for the 2T mode (see Fig.~\ref{r2b}a). For other modes, $\langle\phi\rangle_{p',I'}$ is far off from \ang{0} and also the common spectral power is diminished. Therefore, we observe a weaker coupling for these modes. From these observations, we can surmise that the higher harmonics in $p'$ beyond the first few modes are solely due to the wave steepening effects \cite{saenger1960periodic,chester1964resonant} and do not arise from coupling between the acoustic and the heat release rate oscillations.

\subsubsection{Recurrence analysis of acoustic pressure and \textcolor{black}{CH* intensity oscillations}}
\label{Sec:rec-tai}
The temporal dynamics of a measured signal can be understood by tracking its recurrence in a certain neighborhood in the phase space \cite{marwan2007recurrence}. The recurrence plot (RP) allows one to visually identify the time instants at which the trajectory of the system visits roughly the same region in its phase space \cite{marwan2007recurrence}. The pattern in a recurrence plot enables us to quantify the temporal dynamics of chaotic, quasiperiodic, intermittent, periodic, and stochastic signals \cite{webber2015recurrence}. Before constructing the recurrence plot, we need to compute the phase space of the system. Since an experiment provides only a handful of measurements from the usually high-dimensional system, we need to reconstruct its phase space from the time series of a single measured variable. After computing the optimum time delay ($\tau_{opt}$) and minimum embedding dimension ($d$) of the signal \cite{abarbanel2012analysis}, we follow Takens' delay embedding theorem \cite{takens1981detecting} to reconstruct the phase space. 

For a time series of $n$ time instants, we can obtain the phase space trajectory $\vec{X}(t)$ made of $n - (d-1)\tau_{opt}$ time instants. Then, the pairwise distances between all state points in the phase space can be accommodated in a distance matrix ($D_{ij}$), as formulated below,
\begin{equation}\label{eqs2}
D_{ij} = \left \|\vec{X_i} - \vec{X_j}\right\|  \ \ \ i, j = 1, 2, \ldots, n-(d-1)\tau_{opt}.   
\end{equation}
Here, $\left \|\vec{X_i} - \vec{X_j}\right\|$ is the Euclidean distance between the two state points, $i$ and $j$, on the phase space trajectory. Then, the distance matrix is binarized by applying a suitable threshold ($\epsilon$) to obtain the recurrence matrix ($R_{ij}$).
\begin{equation}\label{eqs3}
R_{ij} = \Theta(\epsilon - D_{ij}),  
\end{equation}
where $\Theta$ is the Heaviside step function and $\epsilon$ is the threshold defining the neighborhood around the state point. The threshold ($\epsilon$) can be fixed as a certain fraction of the size of the phase space attractor. Whenever the phase space trajectory recurs within the region defined by the $\epsilon$ - size ball, we mark 1 in the recurrence matrix. Non-recurring points which traverse beyond the $\epsilon$ - size ball are marked by zeroes in the recurrence matrix. In our RP, we visualize values of one and zero by colored and white points, respectively. Thus, the RP is a two-dimensional arrangement of colored and white points that exhibits different patterns based on the underlying dynamics of the system. For a periodic signal of constant amplitude, we obtain uninterrupted equally spaced diagonal lines in the RP. For random signals, we obtain a grainy structure made up by isolated points in the RP. The different dynamical states during the transition to thermoacoustic instability have been distinguished for the same subscale rocket engine combustor used in this study \cite{kasthuri2019dynamical}.

\begin{figure*}[t]
\includegraphics[scale = 0.158] {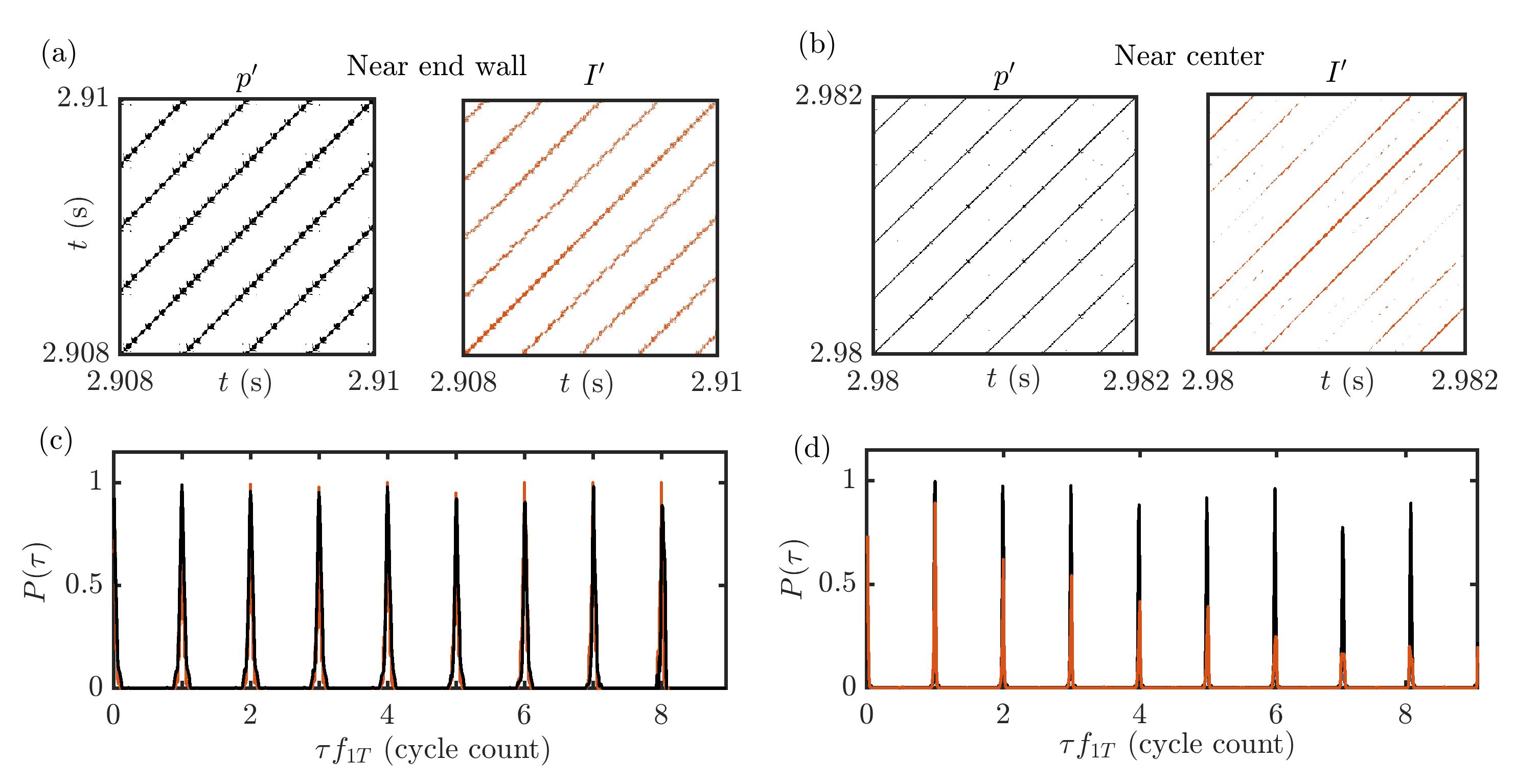}
\caption{The recurrence plots of the acoustic pressure ($p'$) and \textcolor{black}{CH* intensity} fluctuations (\textcolor{black}{$I'$}) during thermoacoustic instability near the (a) end wall and (b) center of the combustor. (c, d) The probability of recurrence $P(\tau)$ is plotted against $\tau f_{1T}$ (i.e., the cycle count) for the end wall and center locations, respectively. The recurrence threshold is fixed as 20\% of the attractor size for each case. }
\label{r4}
\end{figure*}

Following the described methodology, we construct the RP for $p'$ and \textcolor{black}{$I'$} to compare the dynamics during the state of thermoacoustic instability observed from both the locations in the combustor. We employ a threshold of 20\% of the attractor size, and fix the embedding parameters (i.e., the embedding dimension and the time delay) appropriately for each signal. In Fig.~\ref{r4}a,b, we plot the RP for both $p'$ and \textcolor{black}{$I'$} measured near the end wall and center of the combustor, respectively. Near the end wall, the RPs of $p'$ and \textcolor{black}{$I'$} during the state of thermoacoustic instability are nearly identical. Both the RPs are manifested by diagonal lines implying periodicity in the dynamics. Moreover, the RP of $p'$ near the end wall features micropatterns over its diagonal lines as a result of the interplay between slow and fast timescales in the system. For more details on these micropatterns in the RP, readers are directed to \cite{kasthuri2020recurrence}. Such micropatterns are not easily discernible in the RPs of $p'$ and \textcolor{black}{$I'$} measurements near the center of the combustor. Compared to the RPs near the end wall, the RPs of $p'$ and \textcolor{black}{$I'$} near the center of the combustor are not identical. The corresponding RPs can be distinguished by the broken diagonal lines in the RP of \textcolor{black}{$I'$} which are not seen in the RP of $p'$. 

Next, we exploit the quantitative information on the recurrence of $p'$ and \textcolor{black}{$I'$} signals to qualitatively assess their coupled behavior at both locations in the combustor. We compute the probability of recurrence $P$($\tau$) \cite{romano2005detection} which measures the probability with which a trajectory in phase space ($X_i$) revisits the same neighborhood after a time lag $\tau$ and is given as,
\begin{equation}
P(\tau)=\frac{1}{n-(d-1)\tau_{opt}} \sum_{i=1}^{n-(d-1)\tau_{opt}} \Theta(\epsilon-\left\Vert X_i-X_{i+\tau}\right\Vert).
\end{equation}
$P$($\tau$) can be used to capture the type of synchronization existing between $p'$ and \textcolor{black}{$I'$} oscillations from their recurrent behaviors. The type of synchronization is inferred based on the locking of the location of the peaks as well as their heights in the plots of $P(\tau)$ of $p'$ and \textcolor{black}{$I'$} \cite{romano2005detection}. 

Using $P(\tau)$, we can detect the presence of phase synchronization or generalized synchronization amongst the $p'$ and \textcolor{black}{$I'$} signals. During phase synchronization, both the signals show a perfect locking in their instantaneous phases but their instantaneous amplitudes are uncorrelated \cite{romano2005detection}. During generalized synchronization, both the instantaneous phases and amplitudes of the signals are perfectly locked. Therefore, we can express the properties of both signals using a functional relationship (i.e., \textcolor{black}{$I'$} can be modeled as $f(p')$) \cite{romano2005detection}. 

In Fig.~\ref{r4}c,d, we show the plots of $P(\tau)$ calculated for $p'$ and \textcolor{black}{$I'$} as a function of the lag non-dimensionalized by the time period of the cycle (i.e., $\tau f_{1T}$) at both the locations near the end wall and the center of the combustor. Near the end wall, we observe that the locations and the heights of the peaks of $P(\tau)$ for both $p'$ and \textcolor{black}{$I'$} coincide with each other and attain a value close to 1 periodically for each cycle of oscillation. However, near the center of the combustor, $P(\tau)$ of both $p'$ and \textcolor{black}{$I'$} do not have identical heights of their peaks. Moreover, their $P(\tau)$ magnitudes gradually decay for increasing $\tau$. Thus, all these observations in the RPs and the plots of $P(\tau)$ indicate that the dynamics of $p'$ and \textcolor{black}{$I'$} near the end wall is in a state of generalized synchronization. However, near the center of the combustor, we observe a state of phase synchronization between the $p'$ and \textcolor{black}{$I'$} signals. Here, both $p'$ and \textcolor{black}{$I'$} are perfectly phase locked but exhibit a weak correlation between their amplitudes. Thus, the state of generalized synchronization represents a stronger synchronization than phase synchronization since the amplitudes lock to each other in addition to the phase locking between the two signals \cite{romano2005detection, pawar2017thermoacoustic}.

To summarize, we observe a transition from a state of desynchronization (during stable state) to  intermittent phase synchronization (during intermittency) to phase synchronization (during thermoacoustic instability) \textcolor{black}{in the coupled behavior between $p'$ and $I'$ near the center of the combustor}. Due to the stronger coupling between the $p'$ and \textcolor{black}{$I'$} near the end wall of the combustor, we observe generalized synchronization between them. From the mean relative phase information from the XWT plots, we inferred the individual contribution of each of the transverse modes to the coupling between $p'$ and \textcolor{black}{$I'$} near the end wall and the center of the combustor. Near the end wall, the mean relative phase between $p'$ and \textcolor{black}{$I'$} gradually changes from near in-phase for the 1T mode to out-of-phase for the 5T mode. However, near the center of the combustor, we do not observe any such discernible trend in the corresponding mean relative phase.

\subsection{Spatial analysis of the jet flames and their coupled interaction with the transverse acoustics during the transition to thermoacoustic instability}
\begin{figure*}[t!]
\centering
\includegraphics[scale = 0.13] {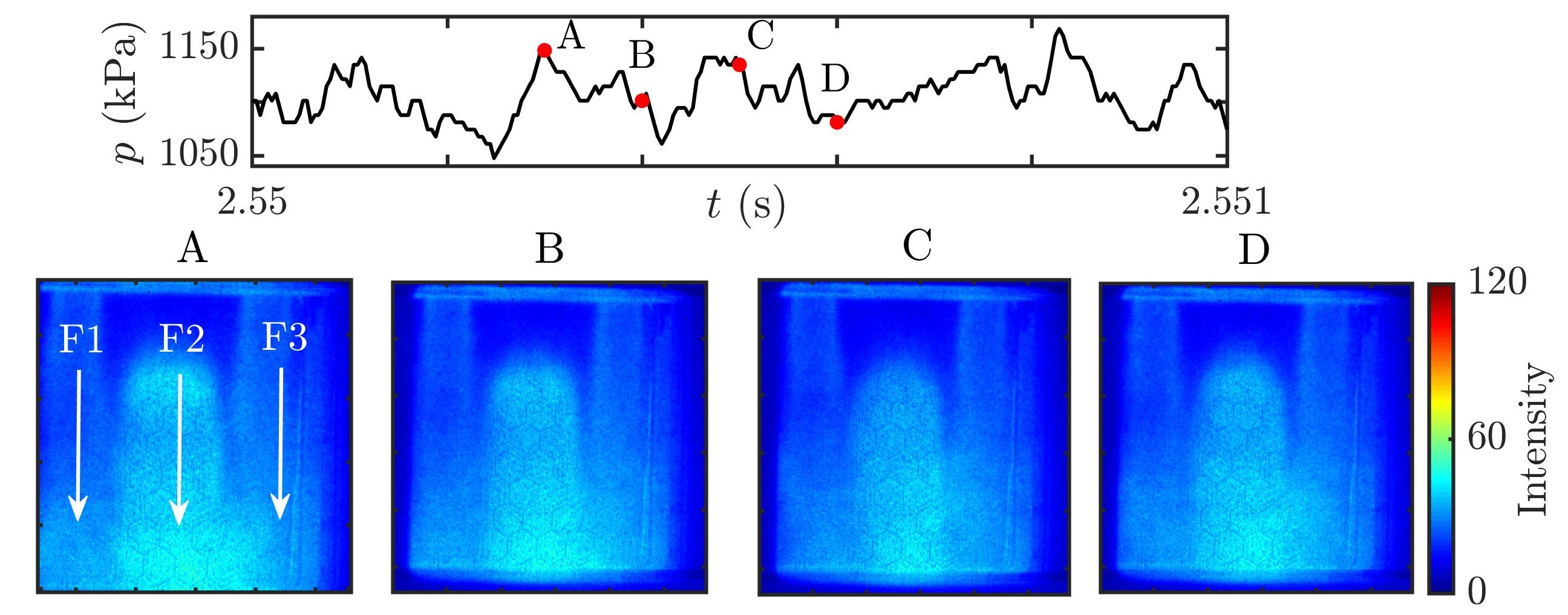}
\caption{Four representative snapshots of the three jet flames at the center of the combustor during stable state. Apart from the central flame, all other flames are anchored to the injector recess. The three flames are represented by arrows in the first instantaneous image (A). }
\label{r5a}
\end{figure*}

Next, we will study the spatiotemporal behavior of the jet flames observed at the two locations in the combustor during the stable state, intermittency, and thermoacoustic instability. The behavior of the jet flames near the center of the combustor during the stable state is represented by four representative instantaneous CH* chemiluminescence snapshots in Fig.~\ref{r5a}. We observe that the jet flames propagate longitudinally in the direction of the flow. The three flames can be clearly distinguished from one another. \textcolor{black}{Only the central flame is lifted off from the injector recess, while its neighboring flames are anchored to the injector recess.} Moreover, the central flame has a shorter jet core and burns rapidly. As a result, we observe higher intensities in the central flame compared to its neighbors. The dynamics of the central lifted flame for low amplitude periodic oscillations during intermittency and high amplitude periodic oscillations during thermoacoustic instability is discussed elaborately with the supporting schlieren images by Gejji et al. \cite{gejji2020combustion}. 

\subsubsection{Phase-averaged flame images during periodic epochs of intermittency and thermoacoustic instability}
\begin{figure*}[t!]
\centering
\includegraphics[scale = 0.135] {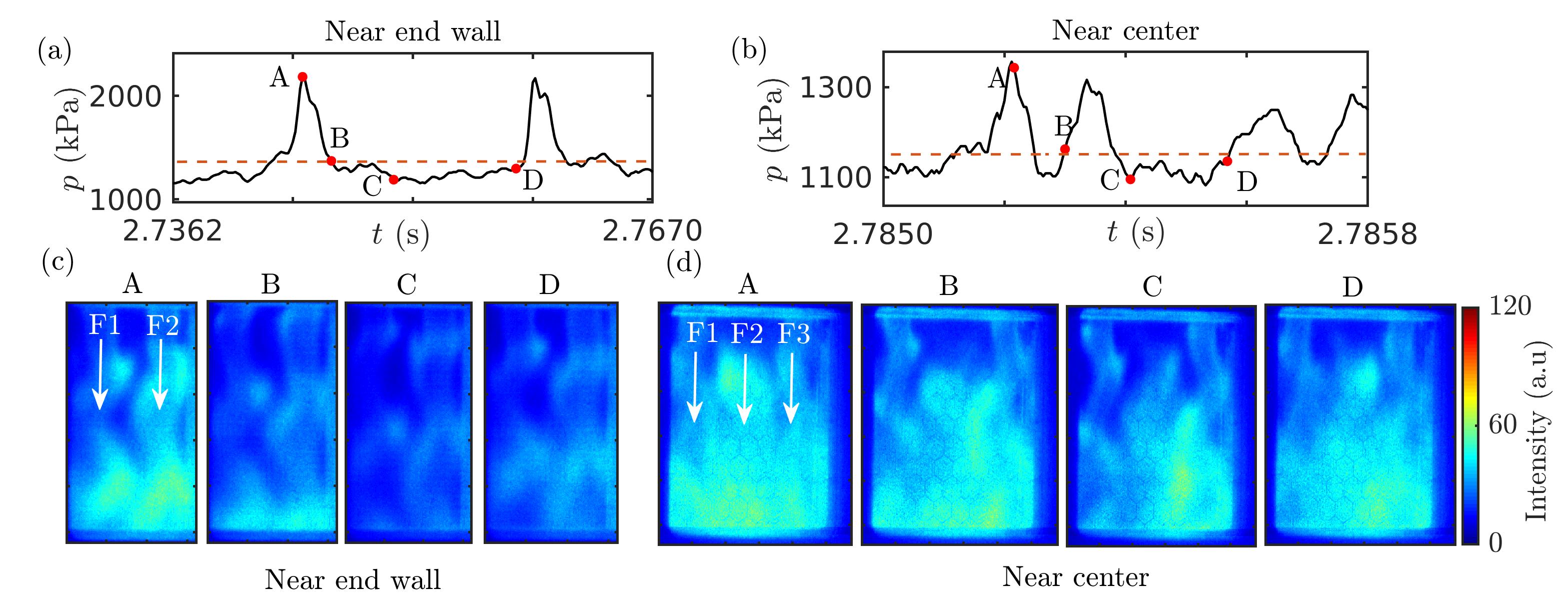}
\caption{Time series of pressure during intermittency at the (a) end wall (PT-01) and (b) the center of the combustor (PT-02). The four phases (A - D) at which the CH* chemiluminescence images are averaged are indicated over the time series. The phase-averaged images during intermittency near (c) the end wall and (d) the center of the combustor are shown. The flames visible through the optically accessible windows are marked in the phase averaged image (at A) for both the locations in the combustor.}
\label{r5ba}
\end{figure*}

The flame behavior during the periodic epochs of intermittency and thermoacoustic instability at the two locations in the combustor are studied by adopting the method of phase averaging. In this method, only the images pertaining to the phase selected for each cycle of oscillations are averaged over 25 cycles. We cannot apply phase averaging during the stable state and for aperiodic epochs of intermittency since the phase for aperiodic oscillations cannot be properly defined. Four phase-averaged CH* chemiluminescence images during periodic parts of intermittency for the two locations in the combustor are shown in Fig.~\ref{r5ba}. The four phases (A - D) indicated over the pressure waveform in Fig.~\ref{r5ba}a are selected to describe the dynamic behavior of the jet flames.

During the periodic epochs of intermittency, at both the locations in the combustor (see Fig.~\ref{r5ba}c,d), we observe that all the jet flames exhibit higher intensities than that observed during the stable state (Fig.~\ref{r5a}). Each jet core remains intact throughout the periodic oscillations observed during intermittency. As a result, the jet flames continue to be distinguishable from each other. Compared to the near steady flames observed during stable state (see Fig.~\ref{r5a}), we observe the presence of periodic transverse displacement of each jet flame during the periodic epochs of intermittency as the steep fronted shock wave sweeps through it. Therefore, we observe that each jet flame exhibits substantial asymmetric oscillations due to the asymmetric vortex shedding from the gaps between the neighboring injector recesses \cite{gejji2020combustion}.

Comparing the phase-averaged images during periodic epoch of intermittency near the end wall (Fig.~\ref{r5ba}c) and center (Fig.~\ref{r5ba}d) of the combustor, we observe that the jet flames near the end wall exhibit higher intensities when the passage of the shock wave coincides with the peak pressure (captured by phase at A). The jet flames near the end wall exhibit their minimum intensity at pressure minima (captured by phase at C) during which the shock wave is far away from the optical window. The images corresponding to phases - B and D show intermediate flame intensities. In contrast to the behavior at the end wall, the shock wave passes through the center of the combustor twice for each reflection off the end wall \cite{harvazinski2019modeling}. Furthermore, the 2T mode is dominant in the center of the combustor, as evidenced from the XWT plot in Fig.~\ref{r2}c. As a result, the jet flames near the center of the combustor  during the periodic part of intermittency exhibit nearly the same intensities for the four phases considered. We also observe that the central jet flame (indicated as F2 in Fig.~\ref{r5ba}d) continues to be lifted off during intermittency. 

\begin{figure*}[t!]
\centering
\includegraphics[scale = 0.137] {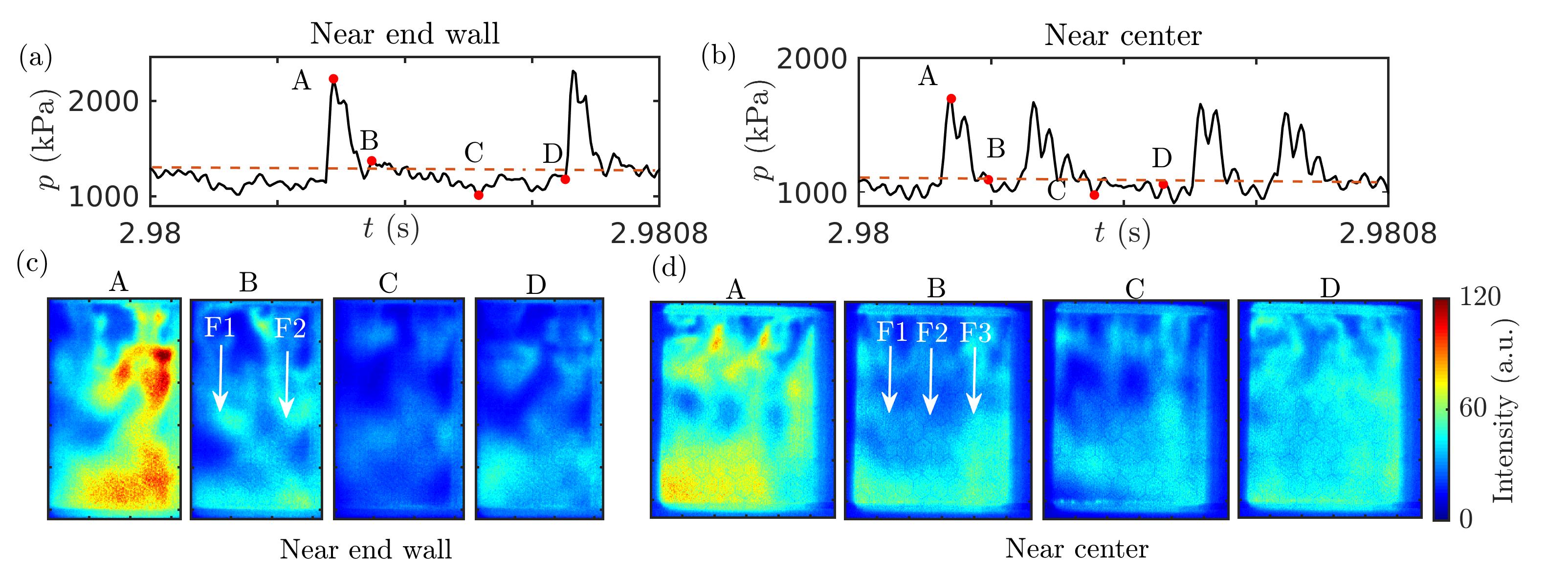}
\caption{Time series of pressure during thermoacoustic instability at the (a) end wall (PT-01) and (b) the center of the combustor (PT-02). The four phases (A - D) at which the CH* chemiluminescence images are indicated over the time series. The phase-averaged images during thermoacoustic instability near (c) the end wall and (d) the center of the combustor are shown. The flames visible through the optically accessible windows are marked in the phase averaged image (at B) for both the locations in the combustor.}
\label{r5bb}
\end{figure*}

In a similar manner, we present the phase-averaged images during thermoacoustic instability at the four phases indicated over the acoustic pressure signal for both the locations in the combustor in Fig.~\ref{r5bb}a,b. Here, we observe significantly higher intensities coinciding with the local pressure maxima (phase A) at both the end wall (Fig.~\ref{r5bb}c) and the center (Fig.~\ref{r5bb}d) of the combustor. Equivalently, the intensities are at their lowest during the pressure minima (phase C). Due to the large transverse oscillations during this state, we observe that the jet cores are no more intact and the jet flames can no more be distinguished from its neighbors. As the shock wave passes through the jet flame, it imparts a large transverse displacement and substantially displaces the jet core, and momentarily results in a spike in the local heat release rate. This spike can be identified from the high intensities observed in the longitudinal location where the jet flames impinge on the end wall (phase - A in Fig.~\ref{r5bb}c). After the shock wave passes through the jet flame, there is a longer relaxation period (phases - B to D). This longer interval allows the fuel and oxidizer to mix and the jet core to regain its original shape. \textcolor{black}{Harvazinski et al. \cite{harvazinski2019modeling} performed hybrid LES/RANS simulations based on the same combustor and operating conditions. From the spatial distribution of methane mass fraction, they found large amounts of methane trapped around the end wall injectors during thermoacoustic instability. The passage of the shock wave rapidly combusts these accumulated reactants, resulting in the enhanced burn rate.} 

As opposed to the flame behavior near the end wall, the continued presence of hot combustion products at the center of the combustor sustains a higher temperature. As a result, there is no excess reactant mixture to be burnt and the jet flames near the center (Fig.~\ref{r5bb}c) are less intense compared to those near the end wall. Moreover, during thermoacoustic instability, the jet flames which are compact near the injector spread out at the downstream locations. Eventually, the propellants auto-ignite leading to higher heat release rate, well downstream of the injector \cite{gejji2020combustion}. 


Thus, during the transition from the stable state to thermoacoustic instability via intermittency, the jet flames exhibit a transition from low intensities to high intensities. This transition is accompanied by a change from a nearly steady longitudinally propagating jet flame during the stable state to a highly unsteady jet flame with significant transverse motion during thermoacoustic instability. The jet flames which are distinct during the periodic part of intermittency exhibit large transverse oscillations during thermoacoustic instability. This results in merging of neighboring jet flames rendering each jet flame indistinguishable from its neighbors. \textcolor{black}{We also established that the jet flames near the end wall exhibit high intensities during the periodic epochs of intermittency and thermoacoustic instability compared to that near the center of the combustor. This difference in the flame behavior across the combustor is attributed to the combination of the end wall housing the pressure antinode and the enhanced combustion from the accumulated unburnt reactants near the end wall.}


\subsubsection{Acoustic power sources and sinks using local Rayleigh index}
Thermoacoustic driving and damping in the combustor can be better understood in terms of acoustic power sources and sinks. The local Rayleigh index ($RI$) is a measure which quantifies the local thermoacoustic driving/damping over a spatial domain \cite{culick2012unsteady}. Since our combustor is receptive to several transverse acoustic modes, we compute the $RI$ for each transverse mode. We define the $RI_{nT}$ for each mode as, 
\begin{equation}\label{eqRI}
RI_{nT} =\frac{1}{TP}\int_{0}^{TP} \frac{p_{nT}'(\vec{x},t)I'_{nT}(\vec{x},t)dt}{\overline{p}_{nT}(\vec{x},t)\overline{I}_{nT}(\vec{x},t)}
\end{equation}
Here, $TP$ is the time period of acoustic pressure oscillation, $nT$ is the $n^{th}$ transverse mode and $\vec{x}$ refers to the location in the two dimensional space of the combustor. The mean of a variable is denoted by overline.

\begin{figure*}[t!]
\centering
\includegraphics[scale = 0.6] {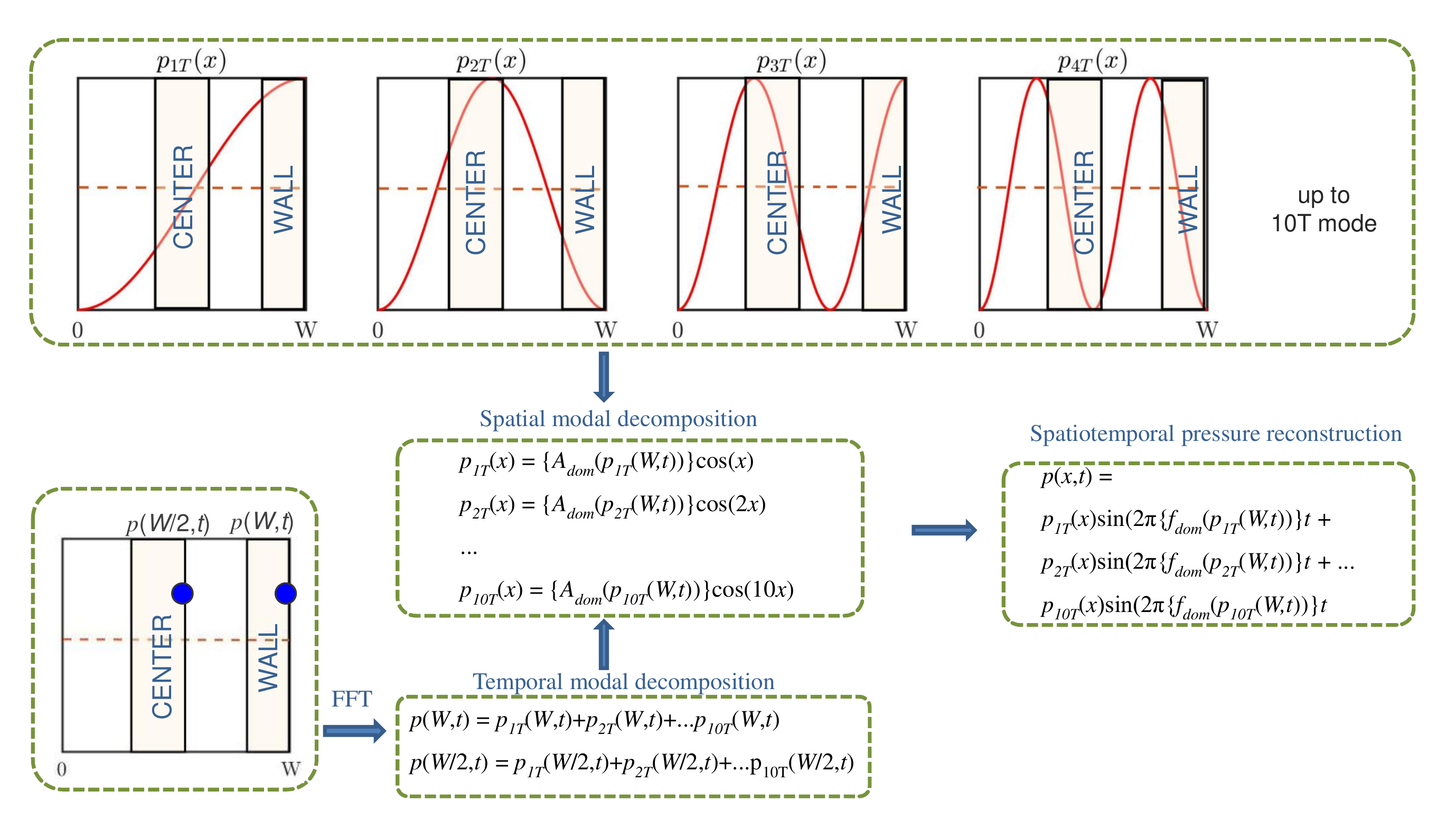}
\caption{Flowchart of spatiotemporal pressure reconstruction based on spatial (only transverse direction) and temporal modal decomposition. The optically accessible windows at the center and end wall of the combustor are shaded for reference. The location of pressure measurements at both these locations are shown using blue circles. Functions $f_{dom}$ and $A_{dom}$ computes the dominant frequency and its corresponding amplitude, respectively.}
\label{r7}
\end{figure*}

The requirement of an unambiguous time period allows us to compute $RI_{nT}$ only during the state of thermoacoustic instability. The local Rayleigh index will allow us to estimate the acoustic driving from each transverse mode in the combustor. However, we need the spatiotemporal pressure and heat release rate variation for each mode. Since the pressure is measured only at specific locations at the center and right side end wall of the combustor (see Fig.~\ref{r0}b), we extract the pressure variation over time and the transverse direction using a spatial and temporal modal decomposition, as depicted in the flowchart shown in Fig.~\ref{r7}. 

\begin{figure*}[t!]
\includegraphics[scale = 0.13] {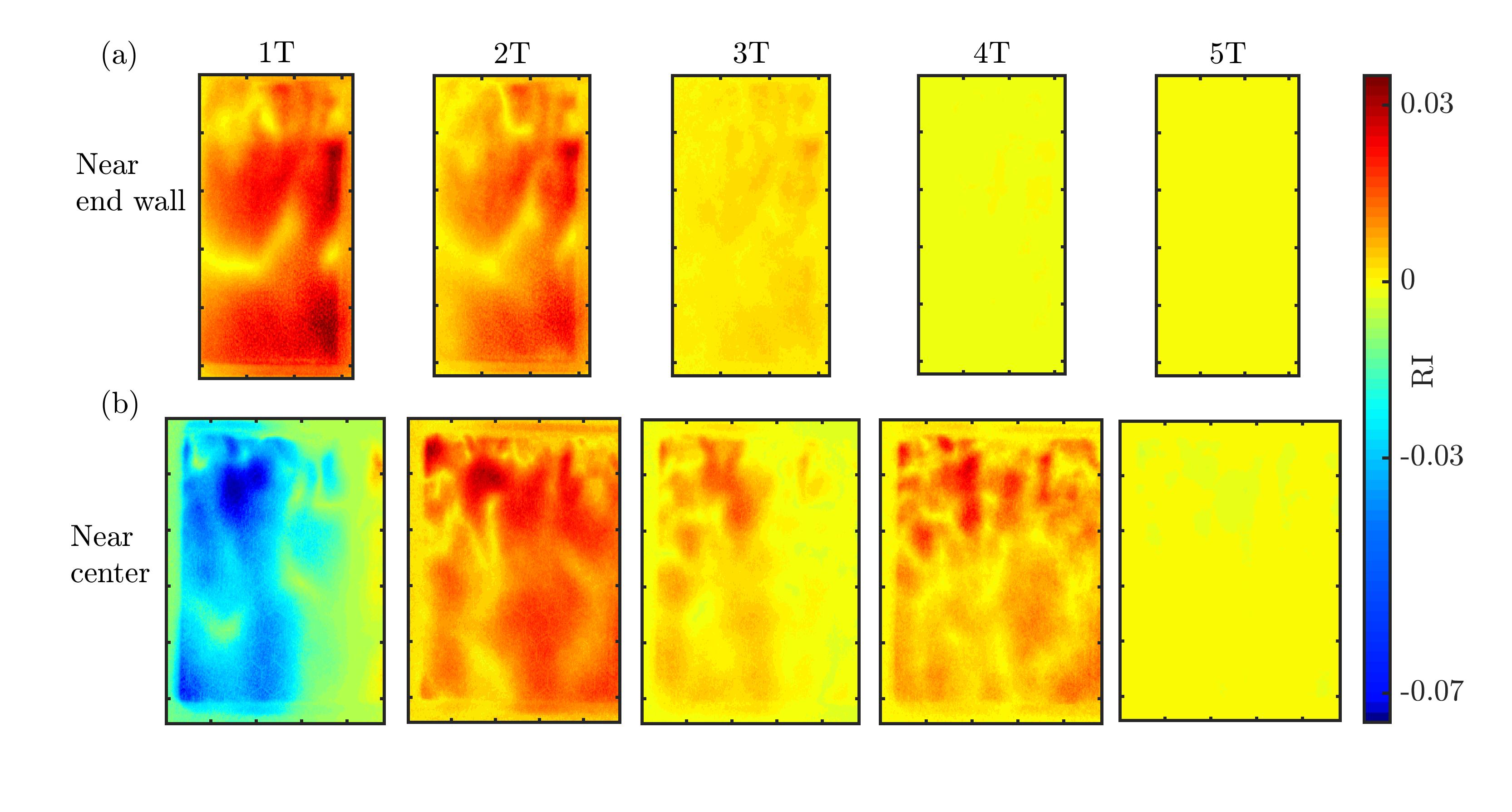}
\caption{The spatial distribution of Rayleigh index ($RI$) is plotted for the first five transverse acoustic modes for near (a) end wall and (b) the center of the combustor during thermoacoustic instability.}
\label{r8}
\end{figure*}

The local \textcolor{black}{CH* intensity oscillations (representative of local heat release rate oscillations)} are decomposed into the individual modes by performing FFT of the CH* chemiluminescence images at each spatial location (i.e., at each pixel). Then, the Rayleigh index for each mode ($RI_{nT}$) is computed over a time interval of 25 cycles of oscillations during thermoacoustic instability. The spatial distribution of $RI_{nT}$ at both the locations considered in this study are presented in Fig.~\ref{r8}. Only the first five acoustic modes are selected since only these modes exhibit significant coupling behavior (as confirmed by the XWT plots in Fig.~\ref{r2}).

Near the end wall (Fig.~\ref{r8}a), we do not observe any significant contribution to acoustic driving beyond the first two modes. We do not observe sufficiently strong acoustic power sources (i.e., $RI\sim$ 0) corresponding to the higher harmonics since these modes are largely out of phase with \textcolor{black}{$I'$} as seen in Fig.~\ref{r2b}b,c. For the 1T and 2T modes, almost the entire window has a uniform distribution of acoustic power sources. Moreover, the distribution of $RI_{1T}$ indicates that the 1T mode provides the highest contribution to acoustic power. Earlier, from Fig.~\ref{r2b}b, we had seen that only the first two modes have $\langle\phi\rangle_{p',I'}$ close to \ang{0}. 

From Fig.~\ref{r8}b, we see that the region near the center derives most of its acoustic power from the 2T mode. We had established that $\langle\phi\rangle_{p',I'}$ at the center of the combustor is closest to \ang{0} only for the 2T mode (Fig.~\ref{r2b}d). In fact, the 1T mode actually houses strong acoustic power sinks that counter the acoustic power contribution from the other harmonics. The damping effect of 1T mode at the center is supported by the anti-phase coupling between $p'$ and \textcolor{black}{$I'$} (see Fig.~\ref{r2b}d). It is interesting to note that the regions occupied by the jet flames offset from the central jet flame in this window has the strongest acoustic sinks (see blue regions in the distribution corresponding to 1T mode in Fig.~\ref{r8}b). The cumulative effect of the contribution of acoustic power sources/sinks from the 1T to 4T acoustic modes is reflected in the weak coupling near the center of the combustor. For thermoacoustic oscillations exhibiting numerous harmonics, we believe this procedure would be useful to quantify the acoustic power sources/sinks from each acoustic mode. 

\subsubsection{Recurrence quantification analysis of local \textcolor{black}{CH* intensity oscillations}}
\begin{figure*}[t!]
\centering
\includegraphics[scale = 0.16] {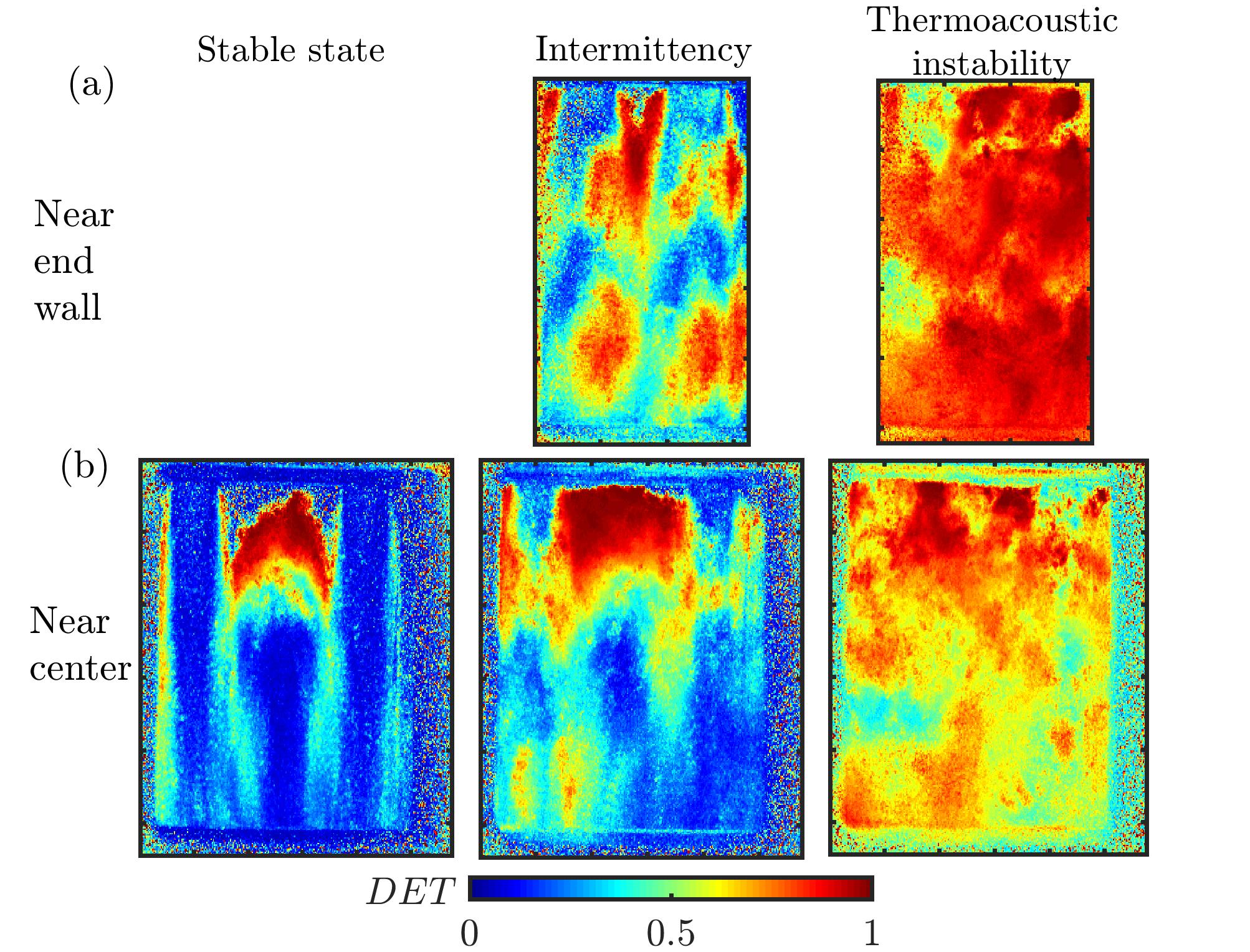}
\caption{The spatial distributions of $DET$ for the local \textcolor{black}{CH* intensity} oscillations measured near the (a) end wall and (b) center of the combustor are plotted for stable state, intermittency, and thermoacoustic instability. A recurrence threshold of 20\% of the attractor size is used to estimate the recurrence measures. Note that stable state is not observed near the end wall of the combustor.}
\label{r6}
\end{figure*}

Next, we perform a recurrence analysis on the local \textcolor{black}{CH* intensity} oscillations observed at both spatial locations of the combustor. Such an analysis would reveal temporal features about the \textcolor{black}{CH* intensity oscillations}, and enable a comparison of the flame dynamics at the two spatial locations. Towards this purpose, we use determinism ($DET$) which quantifies the predictability of \textcolor{black}{CH* intensity} oscillation acquired at a given spatial location. In Fig.~\ref{r6}, we show the spatial distribution of $DET$ for the different dynamical states observed in the combustor. Here, we evaluate $DET$ considering the time series of \textcolor{black}{CH* intensity} oscillations obtained from each pixel within the image. Determinism is derived from the percentage of black points in the corresponding RP which form diagonal lines (representative of periodic behavior) of minimum length $l_{min}$ \cite{marwan2007recurrence}. 

\begin{equation}\label{eqDET}
DET= \frac{\sum_{l=l_{min}}^{n-(d-1)\tau_{opt}} lP(l)} {\sum_{l=1}^{n-(d-1)\tau_{opt}} lP(l)}   
\end{equation}
where $P(l)$ is the probability distribution of diagonal lines having length $l$ in the recurrence plot and $l_{min}$ = 2 \textcolor{black}{time steps}. A purely uncorrelated stochastic signal would have $DET$ extremely close to 0, while a completely deterministic (correlated) signal would have $DET$ = 1. All other signals exhibit $DET$ values between 0 and 1. Hence, $DET$ can be used to distinguish the stochastic or deterministic nature of the signal \cite{marwan2007recurrence}. 

From Fig.~\ref{r6}, we observe that the overall spatial distribution of high values of $DET$ increases as the dynamics evolves from the stable state to thermoacoustic instability. During the stable state, we observe low $DET$ for the spatial locations occupied by the jet cores and higher $DET$ values at the flame edges. This high value of $DET$ arises from the correlated nature of the vortex shedding dynamics along the flame surface. For the central lifted flame (Fig.~\ref{r6}b), we observe that the region between the recess and the flame tip has a very high $DET$, indicating the presence of deterministic fluctuations in the flame anchoring region. Such deterministic behavior in the local \textcolor{black}{CH* intensity} oscillations in the central lifted flame is not easily apparent from the raw CH* chemiluminescence images. During intermittency, with the advent of transverse oscillations, the jet flames get displaced laterally. Correspondingly, we observe patches of high values of $DET$ spread over the window. The locations occupied by the flame just after the entry to the combustor exhibits high $DET$ at both the center and the end wall. Further, we obtain high $DET$ values at locations exhibiting large \textcolor{black}{CH* intensity} oscillations near the end wall (see Fig.~\ref{r5ba}c). As the dynamics transition to thermoacoustic instability, the jet flames are no more intact and spread over the entire window (see Fig.~\ref{r5bb}c,d). Consequently, almost the entire region near the end wall exhibits $DET$ close to 1, indicating widespread periodicity in the local \textcolor{black}{CH* intensity} oscillations. Compared to the end wall (Fig.~\ref{r6}a), the $DET$ near the center (Fig.~\ref{r6}b) varies from 0.5 to 1. This indicates that the \textcolor{black}{CH* intensity oscillations and, in turn, the heat release rate oscillations} during thermoacoustic instability are more deterministic near the end wall than that near the center of the combustor. 

Thus, the spatial distribution of $DET$ is able to capture the deterministic features present in the dynamics of local heat release rate oscillations during each dynamical state. 
\textcolor{black}{Regions exhibiting high/low $DET$ during thermoacoustic instability at the end wall of the combustor coincide with those of high/low $RI_{1T}$, respectively (see the corresponding distribution in Fig.~\ref{r8}a).} However, this \textcolor{black}{qualitative} similarity in the distributions of $DET$ and $RI_{2T}$ is not seen at the center of the combustor.

From this spatial analysis, we have shown that the dynamics of local \textcolor{black}{CH* intensity} oscillations change drastically during the transition from the stable state to intermittency to thermoacoustic instability. Specifically during thermoacoustic instability, the oscillating jet flames in conjunction with the shock wave alters the distribution of the local \textcolor{black}{CH* intensity} oscillations in a significantly different manner at the end wall compared to the center of the combustor. Furthermore, we showed that contribution of local \textcolor{black}{CH* intensity} fluctuations in driving unstable acoustic modes during thermoacoustic instability is significantly dependent on the location within the combustor.

\section{Conclusion}
In this study, we analyzed the coupled interaction between the acoustic pressure oscillations ($p'$) and the \textcolor{black}{CH* intensity} oscillations (\textcolor{black}{$I'$}) in the presence of self-excited transverse thermoacoustic oscillations developed in a multi-element rocket combustor. Specifically, we compared the coupled behavior of these oscillations in the center and the end wall regions of the combustor. During the transition to thermoacoustic instability, we observe a synchronization transition in the coupled behavior of $p'$ and \textcolor{black}{$I'$} oscillations. These oscillations which are desynchronized during stable state and aperiodic epochs of intermittency become phase synchronized during periodic epochs of intermittency. During the state of thermoacoustic instability, we also find that $p'$ and \textcolor{black}{$I'$} exhibit phase synchronization at the center of the combustor and generalized synchronization near the end wall of the combustor. From the increasing trend of the relative mean phase between the modes of $p'$ and \textcolor{black}{$I'$}, we discern that only the first few modes contribute to the coupling between the $p'$ and \textcolor{black}{$I'$} oscillations. The higher harmonics seen in the spectrum of $p'$ arise from the nonlinear wave steepening effect and do not contribute to the coupling between $p'$ and \textcolor{black}{$I'$}. 

Performing a spatial analysis, we found that the local \textcolor{black}{CH* intensity} oscillations near the end wall are higher compared to that near the center of the combustor. This difference in the flame behavior is ascribed to a combination of the presence of pressure antinode and spike in heat release rate due to rapid reaction resulting from wall impingement of premixed pockets of reactants. From the spatial distribution of the Rayleigh index, the contribution of each transverse mode to the generation of acoustic driving is computed during thermoacoustic instability. The superior acoustic driving from the 1T mode near the end wall and the 2T mode near the center of the combustor is revealed. Using recurrence quantification analysis of the local \textcolor{black}{CH* intensity} oscillations, we quantified the transition from stochasticity to widespread determinism in the local \textcolor{black}{CH* intensity} oscillations during the onset of thermoacoustic instability. We also found that the local \textcolor{black}{CH* intensity} oscillations near the end wall are more deterministic (correlated) compared to the center of the combustor during thermoacoustic instability. 

A deeper understanding of the coupling between $p'$ and \textcolor{black}{$I'$} gained from the methods used in this study might be helpful to model the spatial and temporal coupling between the acoustics and the heat release rate fields. Since many real rocket engine combustors experience a combination of longitudinal and transverse modes, it would be interesting to study the coupled interaction and the behavior of local heat release rate oscillations in such configurations. 


\section*{Acknowledgments}
\label{Acknowledgments}
This work was funded by AFOSR under award number: FA2386-18-1-4116 (Grant Program Manager: Lt Col Sheena Winder). R. I. S. thanks Dr. V. Sankaran (AFRL) for initiating this project. We are grateful to Mr. Michael Orth and Prof. Timothee Pourpoint for conducting the experimental campaign at Purdue University.

\newpage
\section*{\textcolor{black}{Appendix A. Desynchrony between acoustic pressure and \textcolor{black}{CH* intensity} oscillations during stable state}}

\begin{figure}[h!]
\centering 
\includegraphics[scale=0.182]{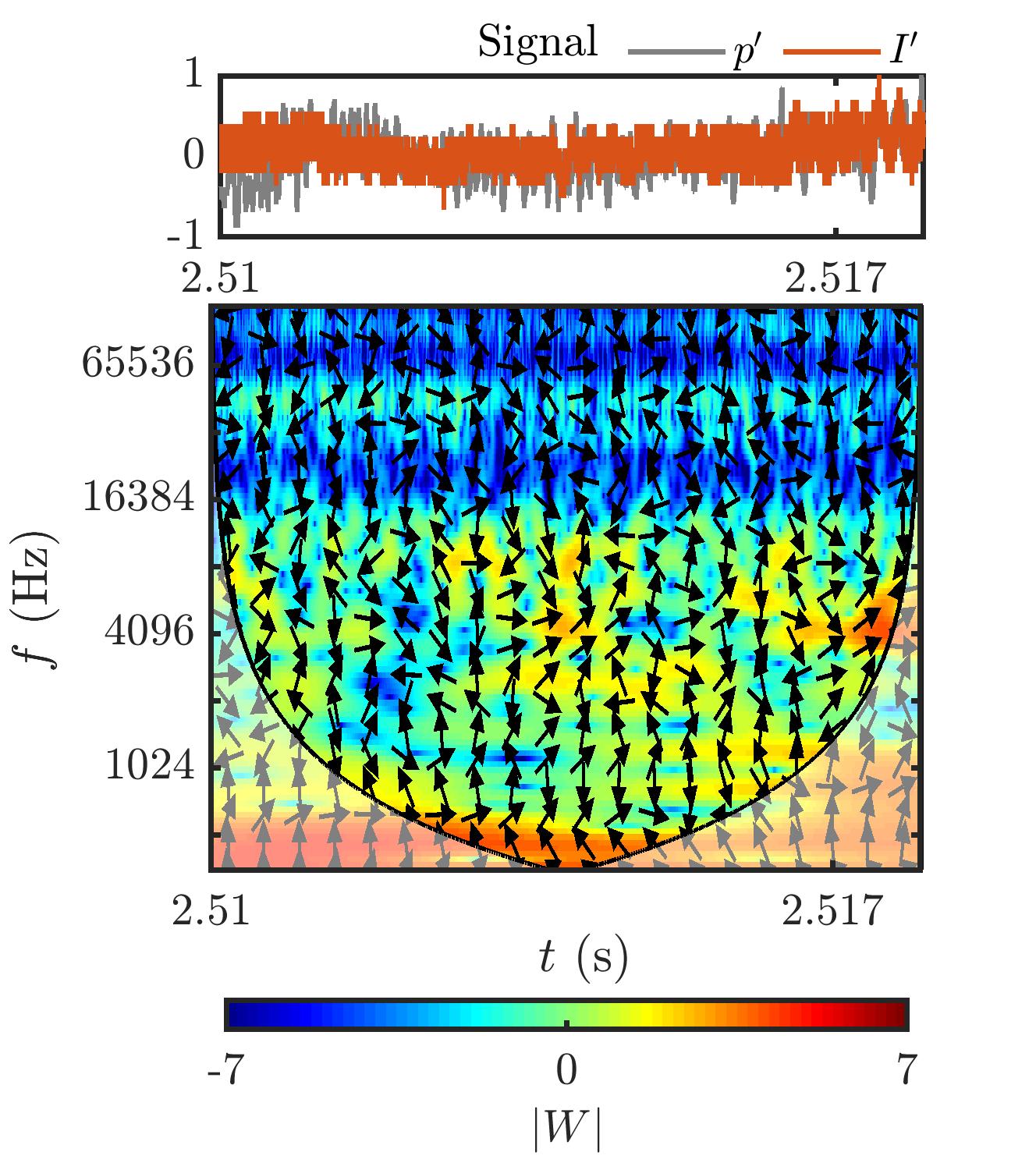}
\caption{\textcolor{black}{The normalized time series of acoustic pressure ($p'$) and CH* intensity ($I'$) oscillations, along with the corresponding XWT plot during the stable state. The pressure and CH* intensity oscillations measured near the center of the combustor are used.}}
\label{fig:XWTsta}
\end{figure}

\textcolor{black}{Both $p'$  and $I'$ are aperiodic throughout the stable state (see Fig.~\ref{r1}I). From the XWT plot of $p'$ and $I'$ shown in Fig.~\ref{fig:XWTsta} for a time interval of 30 ms, we observe a non-homogeneous distribution of common spectral power along with a random alignment of the relative phase arrows at the 1T mode or its harmonics. Therefore, the coupled behavior between $p'$ and $I'$ during stable state corresponds to that of desynchronization.}

\section*{Declaration of competing interest}
The authors declare no competing interests. 

\newpage
\bibliography{mybibfile2.bib} 
\bibliographystyle{elsarticle-num.bst}

\end{document}